\documentclass[prd,nofootinbib,english,superscriptaddress]{revtex4}
\usepackage[utf8]{inputenc}

\usepackage{dsfont}
\usepackage{soul}
\usepackage{comment}
\usepackage{graphicx,color}
\usepackage{rotating}
\usepackage{xcolor}
\usepackage{amssymb}
\usepackage{amsfonts}
\usepackage{amsmath}
\usepackage{cleveref}
\usepackage{xspace}
\usepackage{mathtools} 
\usepackage{braket}
\usepackage{amsmath}

\newcommand\ie{\mbox{\textit{i.\,e.}}\xspace}
\newcommand\cf{\mbox{c.\,f.}\xspace}
\newcommand\eg{\mbox{e.\,g.}\xspace}
\newcommand\D{\mathrm{d}}
\newcommand\hX{\hat{X}}
\newcommand\hV{\hat{V}}

\newcommand\hH{\hat{H}}

\newcommand\hG{\hat{G}}

\newcommand\hx{\hat{x}}
\newcommand\hd{\hat{d}}
\newcommand\hp{\hat{p}}
\newcommand\hk{\hat{k}}

\newcommand\mompar{\dot{\partial}}

\newcommand{\Fabcom}[1]{{\bf\textcolor{orange}{[Fabian: #1]}}}

\begin{document}

\author{Pasquale Bosso}
\email[]{pbosso@unisa.it}
\affiliation{Dipartimento di Ingegneria Industriale, Universit\`a degli Studi di Salerno, Via Giovanni Paolo II, 132 I-84084 Fisciano (SA), Italy}
\affiliation{INFN, Sezione di Napoli, Gruppo collegato di Salerno, Via Giovanni Paolo II, 132 I-84084 Fisciano (SA), Italy}

\author{Giuseppe Fabiano}
\email[]{giuseppe.fabiano@unina.it}
\affiliation{Dipartimento di Fisica Ettore Pancini, Universit\`a di Napoli ``Federico II''}
\affiliation{INFN, Sezione di Napoli, complesso Univ. Monte S. Angelo, I-80126 Napoli, Italy}

\author{Domenico Frattulillo}
\email[]{domenico.frattulillo@unina.it}
\affiliation{Dipartimento di Fisica Ettore Pancini, Universit\`a di Napoli ``Federico II''}
\affiliation{INFN, Sezione di Napoli, complesso Univ. Monte S. Angelo, I-80126 Napoli, Italy}

\author{Fabian Wagner}
\email[]{fwagner@unisa.it}
\affiliation{Dipartimento di Ingegneria Industriale, Universit\`a degli Studi di Salerno, Via Giovanni Paolo II, 132 I-84084 Fisciano (SA), Italy}
\affiliation{INFN, Sezione di Napoli, Gruppo collegato di Salerno, Via Giovanni Paolo II, 132 I-84084 Fisciano (SA), Italy}

\title{The fate of Galilean relativity in minimal-length theories}
\begin{abstract}
    A number of arguments at the interplay of general relativity and quantum theory suggest an operational limit to spatial resolution, conventionally modelled as a generalized uncertainty principle (GUP).   Recently, it has been demonstrated that the dynamics postulated as a part of these models are only loosely related to the existence of the minimal-length scale. In this paper, we intend to make a more informed choice on the Hamiltonian by demanding, among other properties, that the model be invariant under (possibly) deformed Galilean transformations in one dimension. In this vein, we study a two-particle system with general interaction potential under the condition that the composition as well as the action of Galilean boosts on wave numbers be deformed so as to comply with the cut-off. We find that the customary GUP-Hamiltonian does not allow for invariance under (any kind of) generalised Galilean transformations. Those Hamiltonians which allow for a deformed relativity principle have to be related to the ordinary Galilean ones by virtue of a momentum-space diffeomorphism, \ie a canonical transformation. Far from being trivial, the resulting dynamics is deformed, as we show at the example of the harmonic interaction.
    
\end{abstract}
\maketitle

More than a hundred years after its conception \cite{Einstein:1916cbn}, a consistent formulation of a quantum theory of gravity remains elusive (see \cite{Loll:2022ibq} for a recent review). The main reason for this slow progress lies in the scarcity of experimental input. However, recent advances in precision measurements \cite{Amelino-Camelia:1997ieq} as well as control over quantum phenomena \cite{Marletto:2017kzi,Bose:2017nin} have raised hopes that this may change in the near future, leading to the advent of quantum gravity phenomenology \cite{Amelino-Camelia:1999hpv,Amelino-Camelia:2008aez,Addazi:2021xuf}.

One of the main lines of research in quantum gravity phenomenology consists of minimal-length models. As a matter of fact, arguments heuristically combining general relativity and quantum theory suggest the appearance of some kind of minimal-length scale \cite{Mead:1964zz,Padmanabhan:1987au,Maggiore:1993rv,Garay:1994en,Ng:1993jb,Scardigli:1999jh,Adler:1999bu,Calmet:2004mp,Susskind:2005js}. For example, this happens in scattering processes with high center-of-mass energy at impact parameters small enough to create black holes, making it impossible to resolve smaller distances \cite{Scardigli:1999jh,Susskind:2005js,Mead:1964zz}. This intuition is corroborated by circumstantial evidence from explicit approaches to quantum gravity such as string theory \cite{Gross:1987ar,Gross:1987kza,Amati:1987wq,Amati:1988tn,Konishi:1989wk}, loop quantum gravity \cite{Rovelli:1994ge,Modesto:2008jz}, asymptotic safety \cite{Ferrero:2022hor,Lauscher:2005qz}, causal dynamical triangulations \cite{Ambjorn:2005db,Percacci:2010af,Coumbe:2015aev} and Ho\v{r}ava-Lifshitz gravity \cite{Myung:2009gv,Myung:2009ur} (for an extensive review of those motivations see chapter 3 of \cite{Hossenfelder:2012jw}).

In the context of nonrelativistic single-particle quantum mechanics it is customary to introduce the minimal-length scale by deforming the Heisenberg algebra leading to a generalized uncertainty principle (GUP) \cite{Kempf:1994su,Kempf:1996fz,Brau:1999uv,Das:2008kaa,Pedram:2011gw,Nozari:2012gd,Scardigli:2014qka,Bosso:2018uus,Buoninfante:2019fwr,Bosso:2020aqm,Petruzziello:2020wkd,Wagner:2021bqz,Bosso:2022vlz,Wagner:2022rjg,Bosso:2022ogb} (see \cite{Hossenfelder:2012jw,Tawfik:2014zca} and section 3 of \cite{Wagner:2022bcy} for recent reviews, and \cite{Bosso:2023aht} for some critical reflections on the state of the field).
Consequently, the minimal-length scale enters these models by virtue of the Robertson-Schr\"odinger relation \cite{Robertson:1929zz,Schroedinger:1930awq}, \ie as a fundamental limit to localisation. In one dimension, a general parity-invariant modified canonical commutator reads
\begin{equation}
    [\hx,\hp]=if(|\hp|),\label{eqn:DefHeis}
\end{equation}
with the position $\hx$ and the momentum $\hp.$ Depending on the function $f(|\hp|),$ the Robertson-Schr\"odinger relation \cite{Robertson:1929zz,Schroedinger:1930awq}
\begin{equation}
    \Delta x\geq\frac{|\braket{f(|\hp|)}|}{2\Delta p}\label{SchroRo}
\end{equation}
may imply a global minimum to the standard deviation of the position operator.
This is the case, for instance, for the foundational model introduced by Kempf, Mangano, and Mann \cite{Kempf:1994su}
\begin{equation}
    f=1+\ell^2\hp^2,\label{eqn:KMM}
\end{equation}
with the length scale $\ell,$ expected to be of the order of the Planck length.
Equation \eqref{SchroRo} then implies
\begin{equation}
    \Delta x\geq \ell.\label{MinimalLength}
\end{equation}
In short, we choose a function $f$ such that the underlying model exhibits a minimal length. Here, rather than having built a model constructively on the basis of the existence of a minimal length, \ie from the bottom up, we have started by proposing a model, and subsequently shown that it exhibits a minimal length. Top-down approaches of this kind can be instructive when there is an intuition on the choice of model. Unfortunately, in minimal-length quantum mechanics this is not the case. This raises the questions: what is the essence of the minimal length? and which r\^ole shall the function $f$ play from a physical point of view?

Recent developments have marked a step towards solving this puzzle \cite{Bosso:2023sxr,Abdelkhalek:2016nyn,Segreto:2022clx}. If there is to be a minimal length, the kinematics of the theory has to satisfy specific conditions:
given a position operator $\hx,$ we can define its wave-number conjugate $\hk$ such that\footnote{
    Throughout this paper, we will differentiate the terms ``wave number'' and ``momentum'' standing for the generally distinct operators $\hk$ and $\hp,$ respectively.
    Similarly, the terms ``wave-number representation'' and ``wave-number space'' refer to the basis carved out by the eigenstates of $\hk.$}
\begin{equation}\label{heisenberg}
    [\hx,\hk]=i.
\end{equation}
If the standard deviation of the position operator exhibits a global minimum, the spectrum of the operator $\hk$ is necessarily bounded as 
\begin{equation}
    \text{spec}(\hk)=\{k:k\in [-\pi/2\ell,\pi/2\ell]\}.\label{BoundSpecMinLen}
\end{equation}
The constant $\ell$ quantifies the minimal length in the sense that the underlying model obeys Eq. \eqref{MinimalLength}.

In formulating this necessary and sufficient condition for the existence of a minimal length, it has not been necessary to refer to the momentum $\hp$. 
Thus, at first sight it appears that the choice of physical momentum is arbitrary and, most importantly, largely independent of the existence of a minimal length. While this arbitrariness is irrelevant at the kinematic level, it becomes problematic once a Hamiltonian is defined as
\begin{equation}
    \hH=\frac{\hp^2}{2m}+V(\hx),\label{GUPHam}
\end{equation}
\ie in terms of the  momentum $\hp$, as is commonplace in the literature \cite{Kempf:1994su,Kempf:1996fz,Brau:1999uv,Chang:2001kn,Akhoury:2003kc,Benczik:2005bh,Nozari:2005mr,Brau:2006ca,Das:2008kaa,Bouaziz:2008wxq,Das:2009hs,Nozari:2010qy,Bouaziz:2010he,Pedram:2010hx,Pedram:2010zz,Pedram:2011xj,Pedram:2012my,Blado:2013cqb,Das:2014bba,Dey:2015gda,Bosso:2016frs,Bosso:2017ndq,Park:2020zom,Petruzziello:2020een,Bosso:2021koi}.
This particular Hamiltonian is not only not implied by the existence of the minimal length; on the face of it, both are entirely unrelated.
This observation raises two questions \cite{Bosso:2023sxr,Bosso:2023aht}:
if it is not required for the existence of a minimal length, why should we introduce a notion of momentum $\hp$ distinct from the wave number $\hk$ in the first place? and how could we make a more informed guess on the minimal-length deformed Hamiltonian? As we will show below, an answer to the second of these questions entails an answer to the first.

In relativistic theories with an invariant length scale, the choice of physical momentum and its underlying composition law has been addressed in multiple studies \cite{Kowalski-Glikman:2002iba,Amelino-Camelia:2002uql}.
In particular, nonlinearities in the addition law for physical momenta has been the center of the much debated soccer-ball problem \cite{Hossenfelder:2007fy,Hossenfelder:2014ifa,Amelino-Camelia:2014gga}, according to which small Planck-scale modifications
may give rise to drastic macroscopic effects -- in contrast with everyday observations.
As discussed in \cite{Amelino-Camelia:2023rkg}, an unambiguous definition of total physical momentum is only viable when interactions between particles are involved.

In this paper, we derive a unique class of interacting Hamiltonians for two-particle systems in one spatial dimension based on a number of elementary axioms.
Most importantly, we demand that the space of wave numbers be bounded as in Eq. \eqref{BoundSpecMinLen} and that there not be a preferred point, nor a preferred frame.
In other words, we assume the system at hand to be invariant under generalised Galilean transformations, while implying a minimal length of the kind provided in Eq. \eqref{MinimalLength}.
 
In order for wave-number space to be bounded, the addition of wave numbers must cease being linear.
Furthermore, the action of boosts on the wave number (in the ordinary theory a simple translation) has to saturate when approaching the bound. Otherwise this bound could be overshot, for example, in scattering processes or by considering the system from the point of view of a strongly boosted observer.
We find that generalised Galilean invariance of the Hamiltonian, under rather mild assumptions, tightly constrains this composition law, forcing it to be commutative as well as associative. In other words, there must be an operator $\hp=p(\hk),$ which adds up linearly just as momenta in the ordinary theory do, and is, therefore, unbounded. It is this function which, as a matter of convenience rather than necessity, provides a notion of momentum akin to the one implicitly employed in the . Consequently, the deformed commutator in Eq. \eqref{eqn:DefHeis} becomes the inverse Jacobian $f=\D\hp/\D\hk.$
The transformation $\hk\to\hp$ can be turned into a canonical transformation by also scaling the position with its Jacobian to obtain an operator $\hX$, conjugate to $\hp$. Therefore, the Heisenberg equations of motion are left untouched.

As regards the Hamiltonian, we find that the kinetic term indeed provides a nonrelativistic modified dispersion relation of the kind displayed in Eq. \eqref{GUPHam}.
Interaction potentials, however, have to be modified, thus becoming a function of the operator $\hX$ instead of the position $\hx.$
As the phase-space coordinates $(X,p)$ are canonical, the Hamiltonian is thus canonically related to the ordinary quantum mechanical one. 

In short, the only minimal-length model invariant under any deformed version of Galilean relativity is a diffeomorphism
away from the undeformed theory.
As a corollary, conventional GUP models do not allow for a principle of relativity. That the deformed theory can be mapped into the undeformed one reflects the fact that a one-dimensional wave-number space cannot harbour curvature. Indeed, it bears similarity to special relativity in 1+1-dimensions. In contrast to the higher-dimensional case, the latter theory does not possess a curved velocity space. Therefore, it can be mapped into Galilean relativity 
 \cite{Mandanici:2013daa}.

Even though, this implies that the spectrum of the  modified Hamiltonian is undeformed, the ensuing dynamics is by no means trivial, just as special relativity in 1+1 dimensions is not. It is the position $\hx$ that the physical interpretation of the model hinges on because, as was famously laid out in \cite{Feynman:2010qua}, all quantum mechanical measurements come down to positions measurements.
As we show, boosts, \ie now nonlinear changes in the velocity, change the positions of particles dependent on the boost parameter and their momentum. In other words, we find an effect akin to length contraction in special relativity, just that it generally increases distances at large momentum and for fast-moving observers.
To highlight this fact, we consider the Kempf-Mangano-Mann model \cite{Kempf:1994su} as an explicit example, thus showing that distances increase quadratically with the momentum. Furthermore, we explain how the model recovers ordinary Galilean relativity for coarse-grained measurements.

The paper is organised as follows.
First, in section \ref{sec:GalRel} we introduce the notation, deriving the Hamiltonian governing one-dimensional Galilean relativity.
We turn to deformations of Galilean relativity in Sec. \ref{sec:DefGalRel}.
The results are exemplified by the Kempf-Mangano-Mann model in Sec. \ref{sec:KMM}.
Finally, we summarise our results and conclude in Sec \ref{sec:concl}.

\section{Galilean relativity}\label{sec:GalRel}

Before investigating deformed models, it is instructive to see how the dynamical constraints play out in Galilean relativity.
Here, we intend to describe the dynamics of a system of two interacting particles $A$ and $B$ which are governed by an interacting Hamiltonian
\begin{equation}
    \hH=\hH_{0,AB}+\hV,
\end{equation}
with the sum of the ordinary free-particle Hamiltonians $\hH_{0,AB}$ as well as the potential $\hV.$
While the kinetic term $\hH_{0,AB}$ is fixed for arbitrary systems, the potential is left open. Representing the kind of interaction that is to be considered, it generally depends on a function of the positions. 

In one dimension, the Bargmann algebra is spanned by the generators of boosts $\hG_I$ (here the index $I$ can take the values $A$ and $B$), translations $\hk_I,$ and free-particle time translations $\hH_{0,I}$ such that
\begin{align}
    [\hk_I,\hH_{0,I}]=0,&&[\hG_I,\hk_I]=iM_I,&&[\hG_I,\hH_{0,I}]=i\hk_I,\label{eqn:GalAlg}
\end{align}
with the masses of the respective particles $M_I.$ While the first of these commutators indicates $\hH_{0,I}=\hH_{0,I}(\hk_i),$ the other two essentially imply that
\begin{equation}
    \hH_{0,AB}=\frac{\hk_A^2}{2M_A}+\frac{\hk_B^2}{2M_B}.\label{eqn:GalKinTerm}
\end{equation}
Furthermore, considering the fact that the position $\hx_I$ is the conjugate variable to the wave number, we can make use of the Jacobi identity involving $\hx_I,$ $\hk_I$ and $\hG_I$ to identify the Galilean boost generator with the position as
\begin{equation}
    \hG_I=M_I\hx_I.
\end{equation}
Below, we will be interested in the time-evolved version of the boost generator which can be represented as
\begin{equation}
    \hG_{t,I}=e^{i\hH_0 t}\hG_Ie^{-i\hH_0 t}=M_I\hx_I+\hk_It,
\end{equation}
Yet, there is more to the Bargmann algebra than this representation.

To impose a relativity principle to the dynamical structure spelled above, we require that the modification to the Hamiltonian by the transformations generated by $\hk_{AB}=\hk_A+\hk_B$ and $\hG_{t,AB}=\hG_{t,A}+\hG_{t,B}$ be at most a total derivative. As a remnant of the $O(d)$-symmetry which forms part of Galilean invariance, since $O(1)\simeq Z_2,$ we further demand that the Hamiltonian not change under parity transformations, \ie $\hx_I\to-\hx_I,$ $\hk_I\to-\hk_I$. Note here that the translation and boost generators acting on two particles at once are just linear combinations of the ones acting on single particles. Therefore, the algebra in Eq. \eqref{eqn:GalAlg} trivially extends to multi-particle states. The linearity is lost when Galilean relativity is deformed, which will make this extension far less obvious.
The parity transformation, in any case, acts simultaneously on all positions and wave numbers and, being discrete, does not have a generator which could form a part of a Lie algebra.

The kinetic term, being made up only of momenta, is trivially translation invariant. A boost, in turn, shifts it by no more than a total derivative if and only if the free-particle Hamiltonian satisfies Eq. \eqref{eqn:GalKinTerm}. Moreover, Eq. \eqref{eqn:GalKinTerm} is parity invariant. 

We consider potential functions depending on coordinates $\hx^A,\hx^B$ through a distance function $d(\hx^A,\hx^B)$ and present a simple argument required to constrain its form, whose steps we will also employ in the deformed case. In this vein, Galilean invariance requires that the operator $\hd$ be left unchanged under both boosts and translations, \ie $[\hd,\hk_{AB}]=[\hd,\hG_{t,AB}]=0$, which implies
\begin{equation}
    \hd=\hd\left(\hx_A-\hx_B\right).
\end{equation}
Finally, parity invariance renders the sign of $\hx_A-\hx_B$ meaningless, so that the dependence is actually on $|\hx_A-\hx_B|$ . Therefore, the full Hamiltonian reads 
\begin{equation}
    \hH=\frac{\hk_A^2}{2M_A}+\frac{\hk_B^2}{2M_B}+V(|\hx_A-\hx_B|).
\end{equation}
In Galilean relativity, this derivation is, for the most part, straightforward. In the subsequent section, we will see that its deformed variant harbours some slight complications.

\section{Axiomatic approach to deformed Galilean relativity}\label{sec:DefGalRel}

We aim to establish a coherent theory in one spatial dimension that encompasses both single- and multi-particle dynamics while incorporating a fundamental minimal length. As mentioned in the introduction, this minimal length implies a bound to the allowed eigenvalues of the wave number $\hk$, conjugate to the position $\hx.$ Therefore, the wave number necessarily satisfies a deformed composition law of the form
\begin{equation}
    \hk_A\oplus \hk_B=F(\hk_A,\hk_B).\label{eqn:ModWavecomp}
\end{equation}
We make no further assumptions on the function $F$ other than that it recover the usual linear composition law in the limit of vanishing minimal length, \ie $\lim_{\ell\to 0}F=\hk_A+\hk_B.$

We start our argument by stating that the time evolution of the particles in question is to be generated by a Hamiltonian $\hH$ which is given by the sum of a kinetic term $\hH_{0,AB}$ and a potential $\hV.$ For the resulting dynamics to be consistent, we impose the following requirements.
\begin{itemize}
    \item The model allows for a notion of (possibly) deformed Galilean relativity.
    The laws of physics should be the same for every inertial observer connected by symmetry transformations, \ie translations, boosts, and parity transformations, which reduce to their standard expression in the limit of vanishing minimal length.
    In other words, we introduce the translation and boost generators $\hk_I$ and $\hG_I$, respectively, whose action changes the Hamiltonian at most by a total derivative.
    Also, we demand that the generator of time evolution has to be invariant under the standard parity transformation which acts 
    according to
    $\hx_I\to - \hx_I,$ $\hk_I\to-\hk_I,$  $\hG_I\to-\hG_I$.
    \item The model satisfies Newton's first law, \ie in the absence of external fields and for vanishing potential, the time evolution of the generator of translations $\hk_I$ is trivial. This is the case if $[\hk_I,\hH_{0,I}]=0,$ or $\hH_{0,I}=\hH_{0,I}(\hk_I).$
    In other words, the wave number is to be a conserved charge of free-particle motion.
    \item The model allows for free-particles at all energy scales, \ie in multi-particle systems, $\hH_{0,AB}$ equals a simple sum of the kinetic terms of the involved single particles $\hH_{0,I}$. Corrections to the addition law of the kinetic term would imply that highly energetic particles are necessarily interacting nonlocally, thereby, for instance, making it impossible to consider closed systems.
\end{itemize}
In the following, we explore the implications of these axioms on the dynamics of interacting two-particle systems and the composition law provided in Eq. \eqref{eqn:ModWavecomp}. First, we introduce the deformed Bargmann algebra on the level of single particles to subsequently consider interactions.

\subsection{Single particle}

In Galilean relativity, boosts translate in wave-number space.
However, if this very space is bounded, it is not possible to translate indefinitely. In other words, the existence of a bound in wave-number space is incompatible with the action of the standard boost generator on $\hk_I$.
Consequently, we are forced to consider a deformation of the commutator between the boost generator $\hG_I$ and the wave number $\hk_I,$ which, as deformations in minimal-length models scale with the wave number, assumes the form
\begin{align}
   [\hG_I,\hk_I]=iM_Ig(|\hk_I|)\label{eqn:DefGalAlg}
\end{align}
where $g$ is a dimensionless function, tending to $1$ in the limit of vanishing minimal length and saturating towards the bound of wave-number space in order for it to not be exceeded by highly-boosted observers. It can only depend on the wave number in terms of its absolute value by virtue of parity invariance.

Next, we derive the commutator between the boost operator and the kinetic term of the Hamiltonian. Taking into account Newton's first law, $[\hk_I, \hH_{0,I}] = 0$, we conclude that the Hamiltonian is a function of the wave number, \ie $\hH_{0,I} = \hH_{0,I}(\hk_I)$. As a result, we obtain
\begin{equation}
    [\hG_I,\hH_{0,I}(\hk_I)]=iM_I\hH'_{0,I}(\hk_I)g(|\hk_I|).
\end{equation}
To complete the single-particle description, we again employ the Jacobi identity involving $\hx_I,$ $\hk_I$ and $\hG_I$ to represent the boost generator with the position operator as
\begin{equation}
    \hG_I=\frac{M_I\{\hx_I,g(|\hk_I|)\}}{2}\equiv \frac{\{\hat{\bar{G}}_I,g(|\hk_I|)\}}{2}\,,\qquad \hx_I=\frac{\{\hG_I,g^{-1}(|\hk_I)|\}}{2M_I}\label{eqn:BoostJacobi}
\end{equation}
The anticommutator $\{,\}$ is required to preserve Hermiticity with respect to the trivial measure in wave-number space.
Furthermore, we introduced the operator $\hat{\bar{G}}_I,$ the standard Galilean boost, obtained from $\hG_I$ as $\ell\to0,$ \ie $\hat{\bar{G}}_I=M_I\hx_I$.
The generator of time-dependent boosts, in turn, can be represented as
\begin{equation}
    \hG_{t,I}=M_I\left[\frac{\{\hx_I,g\}}{2}+\hH_{0,I}'gt\right].
\end{equation}
In contrast to standard quantum mechanics, in our deformed Galilean framework the relation between the boost generator $\hG_I$ and the coordinate $\hx_I$ is modified nonlinearly. This has remarkable consequences for the construction of deformed relativistic dynamics for multi-particle systems.

\subsection{Interactions between particles}\label{generalmultiparticle}

In Galilean relativity, the extension of kinematics from one to many particles is immediate due to the linearity of the algebra. Yet, this ceases to be the case for nonlinear generalisations thereof as in Eq. \eqref{eqn:DefGalAlg}. In the present subsection, we study the extension of the deformed algebra to multi-particle states, using a system of two particles $A$ and $B$ as a proxy.

For this choice to be compatible with the composition of two boosts, the commutator in \eqref{eqn:DefGalAlg} has to be reproduced in the multi-particle case, namely
\begin{align}
   [ \hG_A\oplus \hG_B,\hk_A\oplus \hk_B]=i(M_A+M_B)g(|\hk_A\oplus \hk_B|)\label{eqn:DefGalAlg2}
\end{align}
In general, the composition of boosts is a function of boosts, wave numbers and masses. However, contributions to the addition law which are nonlinear in $\hG_I,$ by virtue of dimensional analysis, have to be balanced by inverse powers of $\ell M_I$ ($M_I$ here can be any linear combination of the two masses). As these corrections have to disappear in the limit $\ell\to 0,$ they could only contain inverse powers of the operators $\hG_I,$ rendering them nonanalytic. Furthermore, these inverse power would render it impossible to obtain a boost-independent right-hand side in Eq. \eqref{eqn:DefGalAlg2}. We conclude that the composition of boosts has to be linear in the boosts. Therefore, the most general ansatz reads
\begin{equation}
    \hG_A\oplus \hG_B= a_1(\hk_A,\hk_B)\hG_A+a_2(\hk_A,\hk_B)\hG_B
\end{equation}
for the two dimensionless functions $a_1,a_2$ which reduce to $1$ in the limit $\ell\to 0$. Requiring that \eqref{eqn:DefGalAlg2} holds, the boost composition is uniquely fixed, yielding
\begin{equation}
\label{boostsum}
    \hG_A\oplus \hG_B=\frac{1}{2}\left\{\frac{1}{2}\left(M_A\left\{\hx_A, (\mompar^AF)^{-1}\right\}+M_B\left\{\hx_B,(\mompar^BF)^{-1}\right\}\right),g(|\hk_A\oplus\hk_B|)\right\},
\end{equation}
with the derivatives in wave-number space $\dot\partial_I=\partial/\partial \hk_I$. Equipped with the composition law for the relevant symmetry generators of our deformed Galilean framework, we are ready to lay down the foundations to construct relativistic dynamics in the multi-particle case. Following our axiomatic approach, specifically Newton's first law, relativistic invariance demands that its commutator with the combined boost generator $\hG_{A}\oplus \hG_{B}$ at most produces a total derivative. 

Inspired by standard Galilean relativity, we propose that a potential $\hV,$ which commutes with the total boost and the total wave number, be a function of a generalised notion of distance $\hd$, which, in principle, is a function of all phase space variables,\footnote{In contrast to the Galilean case, we cannot assume the potential to be a function of the position only because, on the basis of this assumption, it could not be rendered invariant under generalised Galilean transformations, while at the same time being compatible with the existence of a minimal length.} namely
\begin{equation}
    \hV=V[\hd(\hx_A,\hk_A,\hx_B,\hk_B)].
\end{equation}
We require that $\hd$ is parity invariant and that in the limit of vanishing minimal length it becomes
\begin{equation}
    \lim_{\ell\to 0}\hd=|\hx_A-\hx_B|.
\end{equation}
According to the axioms laid out above, the operator $\hd$ has to be invariant under translations, \ie
\begin{equation}
    U^\dagger_{\hk_A\oplus\hk_B}(a)\hd(\hx_A,\hk_A,\hx_B,\hk_B)U_{\hk_A\oplus\hk_B}(a)\overset{!}{=}d(\hx_A,\hk_A,\hx_B,\hk_B);\label{eqn:TranslPot}
\end{equation}
and under time-dependent boosts, \ie
\begin{equation}
    U^{t,\dagger}_{\hG_A\oplus\hG_B}(u)\hd(\hx_A,\hk_A,\hx_B,\hk_B)U^t_{\hG_A\oplus\hG_B}(u)\overset{!}{=}d(\hx_A,\hk_A,\hx_B,\hk_B),\label{eqn:BoostPot}
\end{equation}
with the time-evolved finite boost transformation
\begin{equation}
    U^t_{\hG_A\oplus\hG_B}(u)=U_{\hH_{0,AB}}(t)U_{\hG_A\oplus\hG_B}(u)U^\dagger_{\hH_{0,AB}}(t).
\end{equation}
For infinitesimal transformations, these conditions imply
\begin{align}
\label{distcond}    \left[\hG_A\oplus\hG_B,\hd\right]=0,&&\left[\hk_A\oplus\hk_B,\hd\right]=0&&\left[\left[\hG_A\oplus\hG_B,\hH_{0,AB}\right],\hd\right]=0,
\end{align}
where the last equality is obtained by applying the Jacobi-identity involving the operators $\hG_A\oplus\hG_B,$ $\hH_{0,AB}$ and $\hd$.

What could the form of the function $\hd$ be? As it generalises the distance function, we require it to have two properties: it should be homogeneous and linear in the coordinates $\hx_A,\hx_B$, given that we are working in one spatial dimension. 
Thus, up to a constant, the most general translation-invariant ansatz reads
\begin{equation}
\label{distanceansatz}
    \hat{d}=\left|\frac{1}{2}\left\{\frac{1}{2}\left(\left\{\hx_A, (\mompar^AF)^{-1}\right\}\right),h_A(\hk_A\oplus\hk_B)\right\}-\frac{1}{2}\left\{\frac{1}{2}\left(\left\{\hx_B, (\mompar^BF)^{-1}\right\}\right),h_B(\hk_A\oplus\hk_B)\right\}\right|,
\end{equation}
for two dimensionless functions $h_A,h_B$ which reduce to $1$ in the undeformed case. The particular parameterisation employed is useful in the calculations that follow. Indeed, imposing that $\hd$ is invariant under the deformed total translation $\hk_A\oplus \hk_B$, we obtain $h_A(\hk_A,\hk_B)=h_B(\hk_A,\hk_B)\coloneqq h(\hk_A,\hk_B)$. 
Here, we introduce the shorthand notation 
\begin{align}
   \hat{\bar{d}}=&\frac{1}{2}\left(\left\{\hx_A, (\mompar^AF)^{-1}\right\}-\left\{\hx_B,(\mompar^BF)^{-1}\right\}\right),\\
    \hat{\bar{G}}_A\oplus\hat{\bar{G}}_B=&\frac{1}{2}\left(M_A\left\{\hx_A, (\mompar^AF)^{-1}\right\}+M_B\left\{\hx_B,(\mompar^BF)^{-1}\right\}\right).
\end{align} Then, the ansatz for $\hd$ simplifies to
\begin{equation}
\label{distansatz1}
    \hd=\left|\frac{1}{2}\left\{\hat{\bar{d}},h(\hk_A,\hk_B)\right\}\right|,
\end{equation}
By virtue of Eq. \eqref{distcond}, the operator $\hd$ has to satisfy $\left[\hG_A\oplus\hG_B,\hd\right]=0$, which becomes equivalent to

\begin{align}
    [\hG_A\oplus \hG_B,\hd]=&\frac{1}{2}\left\{\frac{1}{2}\{\hat{\bar{d}},\left[\hat{\bar{G}}_A\oplus\hat{\bar{G}}_B,h\right]\}+\frac{1}{2}\left\{ h,\left[\hat{\bar{G}}_A\oplus\hat{\bar{G}}_B,\hat{\bar{d}}\right]\right\}\right\},\\
    =&ig\sum_{I=A,B}M_I\left(\frac{1}{2}\left\{\hat{\bar{d}},\frac{\mompar^I h}{\mompar^IF}\right\}\right)-\frac{ig}{2}(M_A+M_B)\frac{\mompar^A\mompar^BF}{\mompar^BF\mompar^AF}\left\{\hat{\bar{d}},h\right\},\label{eqn:BoostCondPotppp}\\
    \overset{!}{=}&0.
\end{align}
where we have used the fact that $\hd$ commutes with any function of the total translation generator $\hk_A\oplus \hk_B$. This condition can be simplified to read
\begin{equation}
    \sum_{I=A,B}M_I\frac{\mompar^Ih}{\mompar^IF}-(M_A+M_B)\frac{\mompar^A\mompar^BF}{\mompar^BF\mompar^AF}h=0.\label{eqn:BoostCondPotpp}
\end{equation}
As there is no independent mass scale in the theory, the only dependence of the function $h$ on the particle masses can be of the product $M_A/M_B.$ Then, the first term of Eq. \eqref{eqn:BoostCondPotpp} can only have the same mass-dependent prefactor as the second one if the function $h$ depends on wave-numbers solely through their composition $\hk_A\oplus\hk_B$. Furthermore, in order for parity invariance to continue to hold, the distance function has to depend on the absolute value of parity-variable quantities. Therefore $h$ has to be an even function of of the wave-number composition, and the operator $\hd$ finally becomes
\begin{equation}
    \hd=\left| h(\hk_A\oplus\hk_B)\hat{\bar{d}}\right|,
\end{equation}
where we removed the symmetric ordering because every function of the translation generator commutes with the generalised coordinate difference. Furthermore, by Eq. \eqref{eqn:BoostCondPotpp} the function $h$ satisfies the differential equation
\begin{equation}
    h'(\hk_A\oplus\hk_B)=\frac{\mompar^A\mompar^BF}{\mompar^BF\mompar^AF}h(\hk_A\oplus\hk_B).\label{eqn:BoostCondPotp}
\end{equation}
The implications of this condition are twofold. On the one hand, it constrains the space of allowed wave-number compositions $F$. On the other hand, given such a wave-number composition $F,$ it determines the function $h.$ 

First, apart from the factor $\mompar^A\mompar^BF/\mompar^AF\mompar^BF,$ all relevant quantities in Eq. \eqref{eqn:BoostCondPotp} depend on the wave-number composition. Thus, for the two terms appearing in Eq. \eqref{eqn:BoostCondPotp} to cancel out, the underlying function $F$ has to satisfy the constraint
\begin{equation}
    \frac{\mompar^A\mompar^BF}{\mompar^AF\mompar^BF}=\tilde{F}(k_A\oplus k_B)\label{eqn:MomAddProp}
\end{equation}
for some function $\tilde{F}.$ As we demonstrate in Appendix \ref{app:compLaw}, this condition forces the composition of wave numbers to be both commutative and associative. As a result, the wave numbers can be mapped to a set of momenta $\hp_I=p(\hk_I)$ whose composition is linear, for some function $p.$ In other words, there are operators $\hp_I$ such that
\begin{align}
    p(\hk_A\oplus \hk_B)=p(\hk_A)+p(\hk_B)=\hp_A+\hp_B,\qquad\iff\qquad F=p^{-1}\circ (p(\hk_A)+p(\hk_B)).\label{eqn:OrdGUPMod}
\end{align}
Here $p^{-1}$ stands for the inverse function which we denote $p^{-1}=k(p)$. Using this definition of momentum, we then obtain the deformed Heisenberg algebra
\begin{equation}
    [\hx_I,\hp_I]=i\frac{\D\hp_I}{\D\hk_I}\equiv if(\hp_I),
\end{equation}
In short, enforcing the relativity principle suggests the use of the momentum $\hp$ which provides us with a GUP of the form given in Eq. \eqref{eqn:DefHeis}.

Second, in terms of the newly introduced momentum $\hp,$ the differential equation \eqref{eqn:BoostCondPotp} can be solved explicitly to yield
\begin{equation}
    h(k)=\frac{\D k}{\D p}(k)=\frac{1}{f\circ p(k)}.\label{eqn:hf}
\end{equation}
As by parity the function $h$ is even in its argument, so has to be the function $f,$ \ie $f(\hp_I)=f(|\hp_I|).$ For reasons of notational simplicity we will henceforth omit this absolute-value sign. By virtue of Eq. \eqref{eqn:hf}, the function $\hd$ assumes the form 
\begin{equation}
    \hd=\left|\frac{\hat{\bar{d}}}{f\circ p(\hk_A\oplus \hk_B)}\right|.
\end{equation}

We now turn to invariance under time-dependent boosts, \ie the last equality in Eq. \eqref{distcond}, which constrains the kinetic part of the Hamiltonian. For convenience, we here recall the condition to be satisfied, \ie
\begin{equation}
    \left[\hd,\left[\hG_A\oplus\hG_B,\hH_{0,AB}\right]\right]=0.
\end{equation}
Given that the boost composition is linear in the coordinates, the commutator between boost and kinetic part of the Hamiltonian will be a function of the wave-numbers $\hk_A,\hk_B$. Yet, the only possible combination of wave numbers that commutes with $\hd$ is the wave-number composition given by the function $F(\hk_A,\hk_B)$, so that we obtain   
\begin{equation}
    \frac{\left[\hG_A\oplus\hG_B,\hH_0\right]}{g(|\hk_A\oplus\hk_B|)}=\left(M_A\frac{\mompar^A\hH_0}{\mompar^AF}+M_B\frac{\mompar^B\hH_0}{\mompar^BF}\right)=\mathcal{F}(F),
\end{equation}
for some function $\mathcal{F}.$  To solve this equation, we again shift to the momenta $\hp_I$, yielding
\begin{equation}
    \frac{M_A\frac{\partial \hH_0}{\partial\hp_A}+M_B\frac{\partial \hH_0}{\partial\hp_B}}{(f^{-1})'\circ(\hp_A+\hp_B)}=\mathcal{F}\circ f^{-1}\circ(\hp_A+\hp_B).
\end{equation}
In other words, the kinetic term of the Hamiltonian satisfies
\begin{equation}
    M_A\frac{\partial \hH_0}{\partial \hp_A}+M_B\frac{\partial \hH_0}{\partial \hp_B}=\tilde{\mathcal{F}}(\hp_A+\hp_B)\label{eqn:KinHamCond}
\end{equation}
for some function $\tilde{\mathcal{F}}.$ From our postulates we recall that the kinetic term for a system of two particles consists of the sum of two independent kinetic terms, \ie
\begin{equation}
    \hH_{0,AB}=\hH_{0,A}+\hH_{0,B}.
\end{equation}
Hence, the only possible solution to Eq. \eqref{eqn:KinHamCond} is
\begin{equation}
    \hH_0=\frac{\hp_A^2}{2M_A}+\frac{\hp_B^2}{2M_B}.
\end{equation}
Finally, we can write down a two-particle Hamiltonian, which is invariant under the deformed Galilean transformations. In all generality, this Hamiltonian reads
\begin{equation}
    \hH=\frac{p^2(\hk_A)}{2M_A}+\frac{p^2(\hk_B)}{2M_B}+V\left(\left|\frac{\hat{\bar{d}}}{f\circ p(\hk_A\oplus \hk_B)}\right|\right)\label{eqn:ModHam}
\end{equation}
where $V$ can be any well-behaved function of the distance $\hd$. As we have found the class of Hamiltonians which are consistent with both the existence of a minimal length and a relativity principle, we can now compare this result to the ansatz towards minimal-length models which is customary in the field. 

\subsection{Shortcomings of the conventional approach}\label{sec:CustApp}

In this subsection, we specialise Eq. \eqref{eqn:ModHam} to a single-particle scenario subject to a potential. This potential approximates an interaction with a classical probe \ie an object with excessively large mass compared with the dynamical single particle such that backreaction effects can be neglected. In this kind of situation the external source provides a preferred frame, where it is at rest situated in the origin. Specifically, let $M_B\to\infty,$ while $\hk_B,\hx_B\to 0$. In this limit, since particle $B$ is considered to be fixed, Eq. \eqref{eqn:ModHam} reduces to 
\begin{equation}
\label{singlelimit}
    \hH=\frac{\hp(\hk)^2}{2M}+V\left(\left|\frac{\left\{\frac{1}{f(\hp(\hk))},\hx\right\}}{2}\right|\right).
\end{equation}
where we removed the subscript $A$ because there is only one particle left. An example for this kind of procedure is the treatment of the hydrogen atom, where the dynamics of the proton are neglected. The resulting dynamics is clearly different from \eqref{GUPHam}, which is the conventional Hamiltonian employed in the context of the GUP \cite{Kempf:1996fz,Brau:1999uv,Chang:2001kn,Akhoury:2003kc,Benczik:2005bh,Nozari:2005mr,Brau:2006ca,Das:2008kaa,Bouaziz:2008wxq,Das:2009hs,Nozari:2010qy,Bouaziz:2010he,Pedram:2010hx,Pedram:2010zz,Pedram:2011xj,Pedram:2012my,Blado:2013cqb,Das:2014bba,Dey:2015gda,Bosso:2016frs,Bosso:2017ndq,Park:2020zom,Petruzziello:2020een,Bosso:2021koi}. Indeed, the (generally sparse) applications of the GUP to multi-particle dynamics in the literature \cite{Quesne:2009vc,Park:2022jrs} adhere to interaction potentials dependent on linear coordinate differences. Thus, the underlying Hamiltonian comes down to the apparently straight-forward generalisation of Eq. \eqref{GUPHam}, \ie
\begin{equation}
    \hH=\frac{p^2(\hk_A)}{2M_A}+\frac{p^2(\hk_B)}{2M_B}+V\left(\left|\hx_A-\hx_B\right|\right).
\end{equation}
Comparison with Eq. \eqref{eqn:ModHam} demonstrates that the conventional GUP-deformed Hamiltonian does not comply with any relativity principle deriving from the algebra given in Eq. \eqref{eqn:DefGalAlg}.

Note, though, that here we only consider potentials that originate in particle interactions. External potentials, \ie solutions to originally (deformed) Galilean invariant field equations, can generally break symmetries. For instance, every curved geometry derived from Einstein's field equations breaks global Lorentz invariance. In the context of elementary quantum mechanical systems, we find this behaviour, for example, when considering the Landau levels of a charged particle subject to a constant magnetic field, thus breaking the $O(d)$ sector, \ie in this case parity symmetry. At present, a consistent description of field dynamics in the presence of a minimal length is lacking. Therefore, the single-particle potentials induced by external fields cannot be clearly determined at this stage.

In \cite{Bosso:2022rue}, authored by one of the present authors, it has been demonstrated that one-dimensional minimal-length theories are incompatible with Galilean invariance. Here we have generalised this statement: one-dimensional minimal-length theories of customary type (where the potential $V(\hx)$ is employed to approximately describe particle interactions) do not allow for any kind of relativity principle, be it ordinary or deformed. We stress that the entire argument behind this reasoning applies on the level of operators, and thus does not resort to any classical notions which could possibly become problematic in the context of the GUP (for more information see \cite{Casadio:2020rsj}, for a different view see \cite{Bosso:2023aht}).

In a nutshell, deformed models that adhere to a relativity principle introduce a departure from the conventional approach. How, then, do they relate to ordinary quantum theory? This question forms the basis of the subsequent subsection.

\subsection{Map to undeformed quantum mechanics}

The momenta $\hp_I$ are defined in such a way that their composition for multi-particle systems is linear. Furthermore, the kinetic term expressed in terms of those momenta appears undeformed. This raises the question: what happens to the model when expressed in terms of the conjugate variables to the momenta $\hp_I$? In that vein, we introduce the operators $\hX_I$ such that
\begin{equation}
    [\hX_I,\hp_J]=i\delta_{IJ}.\label{eqn:Xconjp}
\end{equation}
As both pairs $(\hx_I,\hk_I)$ and $(\hX_I,\hp_I)$ are canonical, going from one to the other amounts to a canonical transformation. 

Plus, Eq. \eqref{eqn:Xconjp} has the solution $\hX_I=\left\{f^{-1}(\hp_I),\hx_I\right\}.$ In order to understand the implications of this fact, let us, for the moment consider classical differential geometry with the slight twist that we use wave-number space as the base manifold of the cotangent bundle. Then, positions $x_I$ are one forms $x_I\D k_I$ (Einstein's sum convention is not applied here). Therefore, a diffeomorphism in momentum space has to be of the form 
\begin{equation}
    k_I\to p_I=p(k_I)\qquad x_I\to X_I=\frac{\D k}{\D p}(k_I)x_I.\label{eqn:MomDiffClass}
\end{equation} 
The transformation
\begin{equation}
    \hk_I\to \hp_I=p(\hk_I)\qquad \hx_I\to \hX_I=\frac{1}{2}\left\{f^{-1}\circ p(|\hk_I|),\hx_I\right\}=\frac{1}{2}\left\{\frac{\D k}{\D p}(\hk_I),x_I\right\}\label{eqn:MomDiffQuant}
\end{equation}
is just the Weyl-symmetric quantisation of Eq. \eqref{eqn:MomDiffClass}. In other words, the descriptions in terms of the pairs $(\hx_I,\hk_I)$ and $(\hX_I,\hp_I)$ are related by a momentum-space diffeomorphism.

That the sets of operators $(\hx_I,\hk_I)$ and $(\hX_I,\hp_I)$ are related by a canonical transformation is well-known in the field of GUPs. Both correspond to different representations of the underlying deformed algebra \cite{Bosso:2021koi}. It remains to be shown how this transformation changes the model Hamiltonian provided in Eq. \eqref{eqn:ModHam}. 

Re-expressing the distance function $\hd$ in terms of the conjugate pair $(\hp_I,\hX_I),$ we find
\begin{equation}
    \hd=\frac{1}{2}\left|\left\{\frac{1}{f(\hp_A)},\hx_A\right\}-\left\{\frac{1}{f(\hp_B)},\hx_B\right\}\right|\equiv\left|\hX_A-\hX_B\right|.\label{eqn:hdRes}
\end{equation}
Consequently, the Hamiltonian can be written as
\begin{equation}
\label{Hamrem}
    \hH=\frac{\hp_A^2}{2M_A}+\frac{\hp_B^2}{2M_B}+V\left(\left|\hX_A-\hX_B\right|\right),
\end{equation}
which by Eq. \eqref{eqn:Xconjp} is equivalent to the Hamiltonian of ordinary Galilean quantum theory with the twist that the operators $\hX_I$ do not  stand for positions. Thus, in one dimension, the minimal length forces us to reinterpret the dynamical variables, while the underlying algebra stays the same, \ie
\begin{equation}
    [\hp_I,\hH_{0,I}]=0,\qquad[\hG_I,\hp_I]=iM_Ig(\hp)f(\hp),\qquad[\hG_I,\hH_{0,I}]=i\hp_Ig(\hp)f(\hp).\label{eqn:DefGalAlg3}
\end{equation}
In other words, the only minimal-length deformed dynamics in 1 spatial dimension, which is compatible with a relativity principle, parity invariance and a trivial composition of kinetic terms, has to be related to the ordinary formalism by a diffeomorphism in momentum space, \ie a canonical transformation.

This is not to say that the theory is trivial. As we demonstrate in Sec. \ref{sec:KMM} with the example of coupled harmonic oscillators, while the spectrum of the Hamiltonian is equal to ordinary quantum mechanics, the modification to the interpretation of the theory is dramatic. As we will make use of boost transformations to relate position measurements of different inertial observers in our example, we first study the properties of deformed boosts.

\subsection{Deformed boosts}\label{sec:defboost}
In the previous subsections, we have formulated a consistent dynamical framework for models involving a minimal length. Each model depends on the choice of two functions, $F(\hk_A,\hk_B)$ and $g(|\hk|)$. Before moving on to an example involving actual dynamics for a system of two particles, we briefly pause to study the consequences of deformed symmetries on the kinematics of single and two-particle systems. From now on, we will focus on a specific subclass of models, for which, upon choosing the deformed sum of wave numbers (namely the function $F$), we constructively derive the function $g$, guided by the fact that the commutator between boost and momenta should saturate when the eigenvalues of the wave-number $\hk$ approach the cut-off. 

The main idea consists in regarding the sum of a finite wave number $\hk_A$ and an infinitesimal wave number $\hk_B$ as an infinitesimal boost transformation with parameter $\hk_B/M_A$ acting on the wave number $\hk_A$, \ie
\begin{equation}
    F(\hk_A,\hk_B)\simeq \hk_A+\left(\frac{\hk_B}{M_A}\right)M_A\, \mompar^B F(\hk_A,0)=\hk_A-i\left(\frac{\hk_B}{M_A}\right)\left[\hG_A,\hk_A\right]
\end{equation}
From the above, we extract the commutator between boost and wave number
\begin{align}
\label{booostcomm}
    \left[\hG_A,\hk_A\right]=&iM_A \mompar^B F(\hk_A,0)\\
    =&\frac{iM_A}{f(\hp_A)}.\label{eqn:Boostcommp}
\end{align}
Since the function $F$ asymptotes to the maximal wave number $\pi/2\ell$, its first derivatives go to zero at the boundary. This guarantees that the right-hand-side of Eq. \eqref{booostcomm} vanishes in that limit, furthermore constraining $\lim_{\hp\to \infty}f(\hp)\to\infty.$ As all prevailing minimal-length models imply a monotonically increasing function $f,$ this demand is rather weak.

Thus, following the outlined procedure to obtain the generator of boosts $\hG_I$, in general, we find
\begin{equation}
    \hG_I=M_I\hX_I\iff g(|\hk_i|)=\frac{1}{f\circ p(|\hk_I|)}.\label{eqn:BoostEqX}
\end{equation}
With this specific choice for $g$, the deformed boost sum in \eqref{boostsum} is entirely specified by $F$, yielding
\begin{equation}
    \hG_A\oplus \hG_B=\frac{\left\{\frac{1}{f (\hp_A)},M_A\hx_A\right\}+\left\{\frac{1}{f (\hp_B)},M_B\hx_B\right\}}{2}=M_A\hX_A+M_B\hX_B.
\end{equation}
As boosts add up linearly and by Eq. \eqref{eqn:BoostEqX}, the operator $\hX_I(\hx_I,\hp_I)$ is invariant under boosts, \ie 
\begin{equation}
    \hX_I(\hx_I',\hp_I')=\hX_I(\hx_I,\hp_I),
\end{equation}
where the primes indicate the boosted quantities. In other words, at equal time a Galilean boost changes the position of any of the two particles as
\begin{equation}
    \hx_I'=U^\dagger_{G_A\oplus G_B}(v)\hx_IU_{G_A\oplus G_B}(v)=\frac{1}{2}\left\{\frac{f(\hp_I+M_I v)}{f(\hp_I)},\hx_I\right\}.\label{eqn:BoostedPosition}
\end{equation}
Recall that by virtue of parity-invariance the function $f$ can only depend on the momentum in terms of its absolute value. Thus, in the boosted frame we can write it as
\begin{equation}
    f(|\hp_I+M_I v|)=f\left(\sqrt{2M_I\hH'_{0,I}}\right),    
\end{equation}
where $\hH'_{0,I}$ denotes the kinetic-energy operator in the boosted frame. In the classical regime,\footnote{Throughout this paper we understand the classical limit as $\hbar\to 0,$ while $\ell/\hbar$ stays constant, a viewpoint which is inherent to the literature on relative locality \cite{Amelino-Camelia:2011lvm}, and has recently been advertised for in the context of the GUP by two of the present authors \cite{Bosso:2023aht}. Otherwise, the classical limit of the GUP is either ill-defined or trivial \cite{Casadio:2020rsj}.} we thus obtain
\begin{equation}
    \braket{\hx_I'}=\frac{f(\braket{|\hp_I+M_I v|})}{f(\braket{|\hp_I|})}\braket{\hx_I}+\mathcal{O}(\hbar)=\frac{f(\sqrt{2M_IE'_{\text{kin}}})}{f(\sqrt{2M_IE_{\text{kin}}})}\braket{\hx_I}+\mathcal{O}(\hbar),
\end{equation}
with the classical kinetic energy $E_{\text{kin}}.$ If, for example, the original description was in the rest frame of particle $I,$ \ie $\braket{\hp_I}=0,$ the boosted position of the particle will be at
\begin{equation}
    \braket{\hx_I'}=f(M_I v)\braket{\hx_I}+\mathcal{O}(\hbar).
\end{equation}
In other words, similarly to special relativity the distance of a particle to the origin changes as a function of the boost parameter $v.$ The difference lies in the fact that the change additionally depends on the original position of the described particle. Thus, for every observer the origin is a preferred point (inasmuch as every object in motion recedes from it). In other words, this property transforms covariantly under translations. Therefore, every observer sees local events unmodified, while distant events change depending on the relative distance and momentum.

Ordinarily, we understand boosts as translations in the space of velocities. In the deformed case, the velocity of a free particle (an observer) reads
\begin{equation}
    \dot\hx_I=-i[\hx_I,\hH_{0,AB}]=f(\hp_I)\frac{\hp_I}{M_I}.
\end{equation}
Therefore, an equal-time boost by $v$ acts on the velocity of a particle as
\begin{equation}
    \dot\hx_I'=f(\hp_I+M_I v)\left(\frac{\dot\hx_I}{f(\hp_I)}+v\right).
\end{equation}
As the function $f$ for conventional minimal-length models is monotonically increasing, this amounts to an additional, possibly nonlinear push if the unboosted momentum is large. In contrast to ordinary Galilean relativity, this push modifies the velocity of distinct particles in different ways which can be inferred from the appearance of their masses and momenta. The relativity principle, however, is unchanged -- observers at different speeds experience the same physics.

\subsection{Deformed translations and relative locality}

We move on to study the effect of total translations on a two-particle system. Let us recall that according to the axioms laid out in Sec. \ref{sec:DefGalRel}, those total translations are generated by the operator $\hk_A\oplus\hk_B$. As usual, let $\hx_I$ denote the position operator of the two particles. By acting with a finite translation on the position operators, we obtain
\begin{equation}
    \hx_I'=U^\dagger_{\hk_A\oplus \hk_B}(a)\hx_IU_{\hk_A\oplus \hk_B}(a)=\hx_I+a \,\mompar_IF({\hk_A\oplus \hk_B})=\hx_I+\frac{f(\hp_I)}{f(\hp_A+\hp_B)}a\label{eqn:TranslPosition1},
\end{equation}
with the translation parameter $a.$ On the classical level, this implies that
\begin{equation}
    \braket{\hx_I'}=\braket{\hx_I}+\frac{f(\braket{\hp_I})}{f(\braket{\hp_A}+\braket{\hp_B})}a+\mathcal{O}(\hbar).\label{eqn:TranslPosition2}
\end{equation}
Consider now these two particles undergoing an elastic collision such that the Heisenberg equations satisfy $\hp_I'(t)\propto\delta(|\hX_A-\hX_B|),$ simulating a classical scattering process. If their expected positions are coincident with the observer's, \ie $\braket{\hx_A}=\braket{\hx_B}=0,$ at least barring quantum corrections, we find that
\begin{equation}
    \braket{\hX_I}=\frac{\braket{\hx_I}}{f(\braket{\hp_I})}+\mathcal{O}(\hbar)=\mathcal{O}(\hbar).
\end{equation}
Thus, if both particles are local to the observer, at lowest order in $\hbar$ the scattering process is indeed taking place locally.

However, if the particles' momenta differ, their positions are not coincident for the translated observer who expects
\begin{equation}
    \braket{\hx_I'}=\frac{f(\braket{\hp_I})}{f(\braket{\hp_A}+\braket{\hp_B})}a+\mathcal{O}(\hbar).
\end{equation}
In other words, to the translated observer, the particles appear to interact nonlocally if their momenta differ in absolute value. Whether the interaction is local, therefore depends on the observer. This is an instance of relative locality \cite{Amelino-Camelia:2011lvm,Amelino-Camelia:2011hjg}. Note, however, that quantum corrections can generally change this conclusion.

We thus conclude our investigation on the consequences of general deformations of the Bargmann algebra. To further highlight the implications of this modification, it is instructive to study a specific example, which we do in the subsequent section.
\section{Case study: Kempf-Mangano-Mann model}\label{sec:KMM}

The classic minimal-length model which continues to be in customary use goes back to Kempf, Mangano and Mann \cite{Kempf:1994su}. As provided in Eq. \eqref{eqn:KMM}, it purports a second-order correction between the position and the momentum operators, \ie
\begin{equation}
    [\hx_I,\hp_I]=i\left(1+\ell^2\hp_I^2\right),
\end{equation}
where $\ell$ again plays the r\^ole of minimal length. The wave-number conjugate to the position $\hx$ introduced here is related to the momentum as
\begin{equation}
    \hp_I=\frac{\tan (\ell\hk_I)}{\ell}.
\end{equation}
Assuming that the momenta of the particles in question are composed linearly, the wave numbers have to obey the deformed addition law
\begin{equation}
        F(\hk_A,\hk_B)=\hk_A\oplus \hk_B=\frac{1}{\ell}\arctan\left(\tan(\ell \hk_A)+\tan(\ell\hk_B)\right).
\end{equation}
Following the argument of section \ref{sec:defboost}, the commutator between boost and wave number reads
\begin{equation}
[\hG_I,\hk_I]=iM_I\cos^2(\ell\hk_I).
\end{equation}
As required, the action of boosts on wave numbers saturates at the boundary of wave-number space such that it cannot be overshot.

The conjugate variables to the momentum operator from which the operator $\hd$ is constructed by Eq. \eqref{eqn:hdRes}, read
\begin{equation}
    \hX_I=\frac{1}{2}\left\{\frac{1}{1+\ell^2\hp_I^2},\hx_I\right\}=\frac{1}{2}\left\{\cos^2 (\hk_I\ell),\hx_I\right\}=\frac{\hG_I}{M_I}.
\end{equation}
Consequently, a boost by a velocity $v$ acts on the semiclassical position of a particle at rest as
\begin{equation}
    \braket{\hx_I'}=\left[1+(\ell M_Iv)^2\right]\braket{\hx_I}+\mathcal{O}(\hbar).
\end{equation}
In other words, the distance from the origin increases with large boosts. Having all required operators in place, we can study the modification to the ordinary Galilean theory in evaluating the expectation value of $\hd$ in typical states of interest.

\subsection{Generalised Gaussian states}\label{sec:GenGauss}

In general minimal-length models, there is no physical position representation because the eigenstates of the position operator, which are infinitely peaked, are not contained in the physical Hilbert space; the latter requires a minimal position uncertainty (\ie Eq. \eqref{MinimalLength}). Instead, it is possible to construct a quasi-position representation \cite{Kempf:1994su} from so-called minimal-uncertainty states. These constitute a generalisation of Gaussian states, defined such that they saturate the Robertson-Schr\"odinger relation \cite{Robertson:1929zz,Schroedinger:1930awq} of $\hx_I$ and $\hp_I,$ \ie the GUP \cite{Kempf:1994su}. Such a minimal-uncertainty state at the average positions $\braket{\hx_I}$ and with vanishing expected momenta reads \cite{Bosso:2020aqm}
\begin{equation}
    \psi_{\braket{x}_I}(k_I)=\frac{\ell}{2\sqrt{\pi}}\prod_{I=A,B}\sqrt{\frac{\Gamma (1+a_I)}{\Gamma (\frac{1}{2}+a_I)}}\cos(\ell k_I)^{a_I}e^{-ik_I\braket{x_I}},\qquad\text{with~}a_I=\frac{1+\ell^2\Delta p_I^2}{2\ell^2\Delta p_I^2},\label{eqn:GenGauss}
\end{equation}
where we introduced the Euler Gamma-function $\Gamma(x).$ The quasi-position representation, made up of states of largest possible localisation (\ie saturating Eq. \eqref{MinimalLength}), is then obtained for $\Delta p=\ell^{-1}$. 

Given such a state, the expectation value of the operator $\hX_I$ becomes
\begin{equation}
    \braket{\hX_I}=\frac{1+2\ell^2\Delta p_I^2}{1+3\ell^2\Delta p_I^2}\braket{\hx_I}.
\end{equation}
In other words, while the expectation values of $\hx_I$ and $\hX_I$ coincide in the limit $\ell\Delta p_I\to 0,$ with increasing momentum uncertainty $\braket{X_I}$ decreases to finally equal $2\braket{\hx_I}/3$ in the limit $\ell\Delta p_I\to\infty.$ For states comprising the quasi-position representation, we obtain
\begin{equation}
    \braket{\hX_I}=\frac{3}{4}\braket{\hx_I},
\end{equation}
which is independent of the minimal length. In other words, strongly localised states imply macroscopic differences to observables. This was to be expected because this amount of localisation requires momenta at the level of the minimal length, \ie exactly $\Delta p_I=\ell^{-1}$.

Most importantly, the expectation value of the operator $\hd^2,$ the argument of the potential in Eq. \eqref{Hamrem}, becomes approximately
\begin{equation}
    \braket{\hd^2}=\left(\braket{x_A}-\braket{x_B}\right)^2+\frac{1}{4\Delta p_A^2}+\frac{1}{4\Delta p_B^2}-2\ell^2(\braket{x_A}-\braket{x_B})\left(\Delta p_A^2\braket{x_A}-\Delta p_B^2\braket{x_B}\right)+\mathcal{O}(\ell^4).
\end{equation}
Consequently, the expectation value of the argument of the potential fulfils the expectation of the expected distance between two Gaussian states in the limit of vanishing minimal length. For the constituent states of the quasi-position representation, however, we obtain exactly
\begin{equation}
   \braket{\hd^2}=\ell^2+\frac{5\left(\braket{\hx_A}^2+\braket{\hx_B}^2\right)-9\braket{\hx_A}\braket{\hx_B}}{8},
\end{equation}
which, independently of the value of $\ell$ yields macroscopic changes to the value of the generalised distance. Thus, for all intends and purposes, no particle has ever been detected in a quasi-position eigenstate.\footnote{By analogy with Lorentzian-relativistic quantum mechanics this can be understood as an argument in favor of using positive operator valued measures \cite{Peres:2002wx} to model measurements instead of simple projections on eigenstates.}

To gain an intuition on the consequences of the modifications analysed in the present section, it is instructive to consider an explicit example.
Therefore, in the following, we analyse the coupled harmonic oscillator.

\subsection{Coupled harmonic oscillator}

We have seen that, in one dimension, the dynamics of every system obeying a deformed version of Galilean relativity can be mapped into ordinary quantum mechanics by virtue of a canonical transformation. In other words, we may implement a minimal length by describing the kinematics in terms of the canonical pair $(\hx,\hk),$ where the spectrum of $\hk$ is bounded. This representation is momentum-diffeomorphically related to the canonically conjugate operators $(\hX,\hp)$ satisfying ordinary Galilean relativistic dynamics. Nevertheless, the resulting model is by no means trivial. In this section, we explore some of the consequences of this construction with the help of a simple yet illustrative example -- the coupled harmonic oscillator.

As we have demonstrated in Sec. \ref{sec:CustApp}, the single-particle Hamiltonian obeying a deformed version of Galilean relativity is given by Eq. \eqref{singlelimit}. Expressed in terms of the pair $(\hp,\hX),$ it thus reads
\begin{equation}
    \hH=\frac{\hp^2}{2M}+V(\hX).\label{eqn:GalHam}
\end{equation}
Hence, the energy eigenspectrum is generally undeformed. However, the dynamics is nontrivial just because the equations of motion for the position $\hx$ are nontrivial.

As a specific system, consider two particles of equal mass $M$ connected by a spring. The resulting Hamiltonian reads
\begin{equation}
    \hH=\frac{\hp_A^2+\hp_B^2}{2M}+\frac{M\omega^2}{4}\left(\hX_A-\hX_B\right)^2,
\end{equation}
with the oscillation frequency $\omega.$ The dynamical equations can be decoupled by dividing the motion in $X$-space into a center-of-mass contribution and a relative part such that
\begin{align}
    \hX_{\rm{com}}=\frac{\hX_A+\hX_B}{2},&&\hp_{\rm{com}}=\hp_A+\hp_B,&&\hX_{\rm{rel}}=\frac{\hX_A-\hX_B}{2},&&\hp_{\rm{rel}}=\hp_A-\hp_B,
\end{align}
which is a canonical transformation. As a result, the Hamiltonian becomes
\begin{equation}
    \hH=\frac{\hp_{\rm{com}}^2+\hp_{\rm{rel}}^2}{2M_{\text{tot}}}+\frac{1}{2}M_{\text{tot}}\omega^2\hX_{\rm{rel}}^2,
\end{equation}
with the total mass $M_{\text{tot}}=2M.$ Thus, the dynamics comes down to a simple harmonic oscillator in $X$-space. Thus $\hp_{\rm{com}}$ and consequently $\hk_{\rm{com}}=k(\hp_{\rm{com}})$ are constants of motion as required by Newton's first law. 

 We are working in the Heisenberg picture such that states stay constant while operators evolve in time according to the Heisenberg equation. For the pairs $(\hX_I,\hp_I)$ we thus obtain
\begin{align}
    \hX_A(t)=&\hX_{\rm com}(0)+\frac{\hp_{\rm{com}}t}{M_{\text{tot}}}+\hX_{\rm{rel}}(0) \cos(\omega t)-\frac{\hat{p}_{\rm rel}(0)}{M_{\text{tot}}\omega}\sin (\omega t)=2\frac{\hp_{\rm{com}}t}{M}-\hX_B(t),\\
    \hp_A(t)=&\frac{1}{2}\left[\hp_{\rm{com}}-M_{\text{tot}}\omega\hX_{\rm{rel}}(0)\sin\left(\bar{\omega} t\right)-\hp_{\rm rel}(0)\cos (\omega t)\right]=\hp_{\rm{com}}-\hp_B(t),
\end{align}
with the operator-valued relative-position operator at the beginning of the evolution $\hX_{\rm rel}(0),$ initial center-of-mass position $\hX_{\rm com}(0)$ and initial relative momentum $\hp_{\rm rel}(0)$. The evolution of the position operators can then be inferred as
\begin{align}
    \hx_I(t)=\hX_I(t)+\frac{1}{2}\left\{\ell^2\hp_I(t)^2,\hX_I(t)\right\}.
\end{align}
Thus, we can express the time-evolution of the position operators in terms of the operators $\hX_{\rm{rel}}(0),$ $\hX_{\rm com} (0),$ $\hp_{\rm rel}(0)$ and $\hp_{\rm{com}}.$ Furthermore, we can apply a deformed Galilean boost (with boost-parameter $v$) to the system by shifting
\begin{align}
    \hX_{\rm{rel}}\to\hX_{\rm{rel}},&&\hp_{\rm{com}}\to \hp_{\rm{com}}+M_{\rm tot}v.
\end{align}

In order to study the evolution of the expected position a typical system exhibits, we consider the generalised Gaussian states defined in Eq. \eqref{eqn:GenGauss}. In the limit of $\Delta p_I\ell\to 0$ these are coherent states, thus closely mimicking classical evolution. As given in Eq. \eqref{eqn:GenGauss} the generalised Gaussian states have vanishing expected momentum for both particles. Thus, initially, the center-of-mass momentum of the system vanishes in the unboosted frame $v=0.$

There are four dimensionless parameters that can indicate strongly deformed evolution when being at least of order one, namely $M\omega\ell^2$ ($M\omega$ constitutes the relevant momentum scale of the oscillator), $\Delta p_A(0)\ell,$ $\Delta p_B(0)\ell$ (precision to which momentum/ position is known initially) and $Mv\ell$ (strength of the boost). 

\begin{figure}
    \centering
    \includegraphics[width=\linewidth]{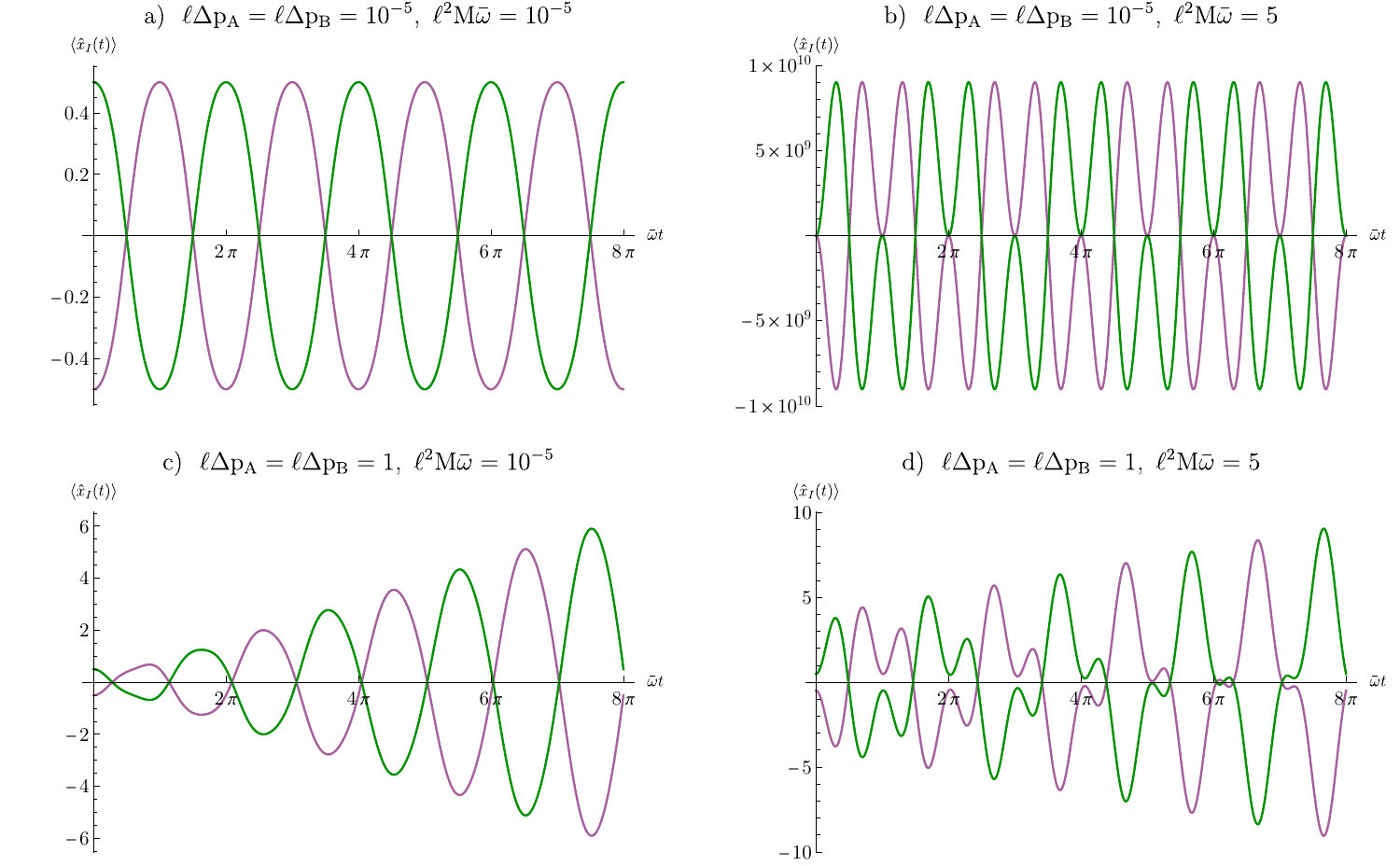}
    \caption{Time evolution of $\braket{\hx_I(t)}$ in units of the coordinate distance between the two particles at $t=0$ with increasing momentum uncertainty from left to right as well as increasing oscillator momentum scale $M\omega$ from top to bottom. The green and violet lines stand for $\braket{\hx_A(t)}$ and $\braket{\hx_B(t)},$ respectively. Additional parameter values are $\braket{\hx_A(0)}=-\braket{\hx_{B}(0)}=\ell/2=1/2$ and $v=0.$ The black dots symbolise the end of a period.}
    \label{fig:HarmOscAllunboosted}
\end{figure}

The ensuing evolution of the expected position is displayed for combinations of the first three parameters in the unboosted stage in Fig. \ref{fig:HarmOscAllunboosted}. In this vein, Fig. \ref{fig:HarmOscAllunboosted} a demonstrates that the evolution is basically undeformed if the relevant parameters are small. An increase in the system-characteristic momentum scale $\sqrt{M\omega}$ induces higher modes of oscillation, overtones of fractional period with respect to $\bar{\omega},$ which leads to the two particles sometimes scattering off each other, while other times simply passing by, clearly very unusual behaviour (see Figs. \ref{fig:HarmOscAllunboosted} b and d). Strong positional localisation, in turn, shifts the phase and frequency of the oscillator while at the same time leading to a constant increase in the separation of the particles (see Figs. \ref{fig:HarmOscAllunboosted} c and d). Note that while those latter plots appear to indicate an instability, the energy along the evolution is constant as expected.

\begin{figure}
    \centering
    \includegraphics[width=\linewidth]{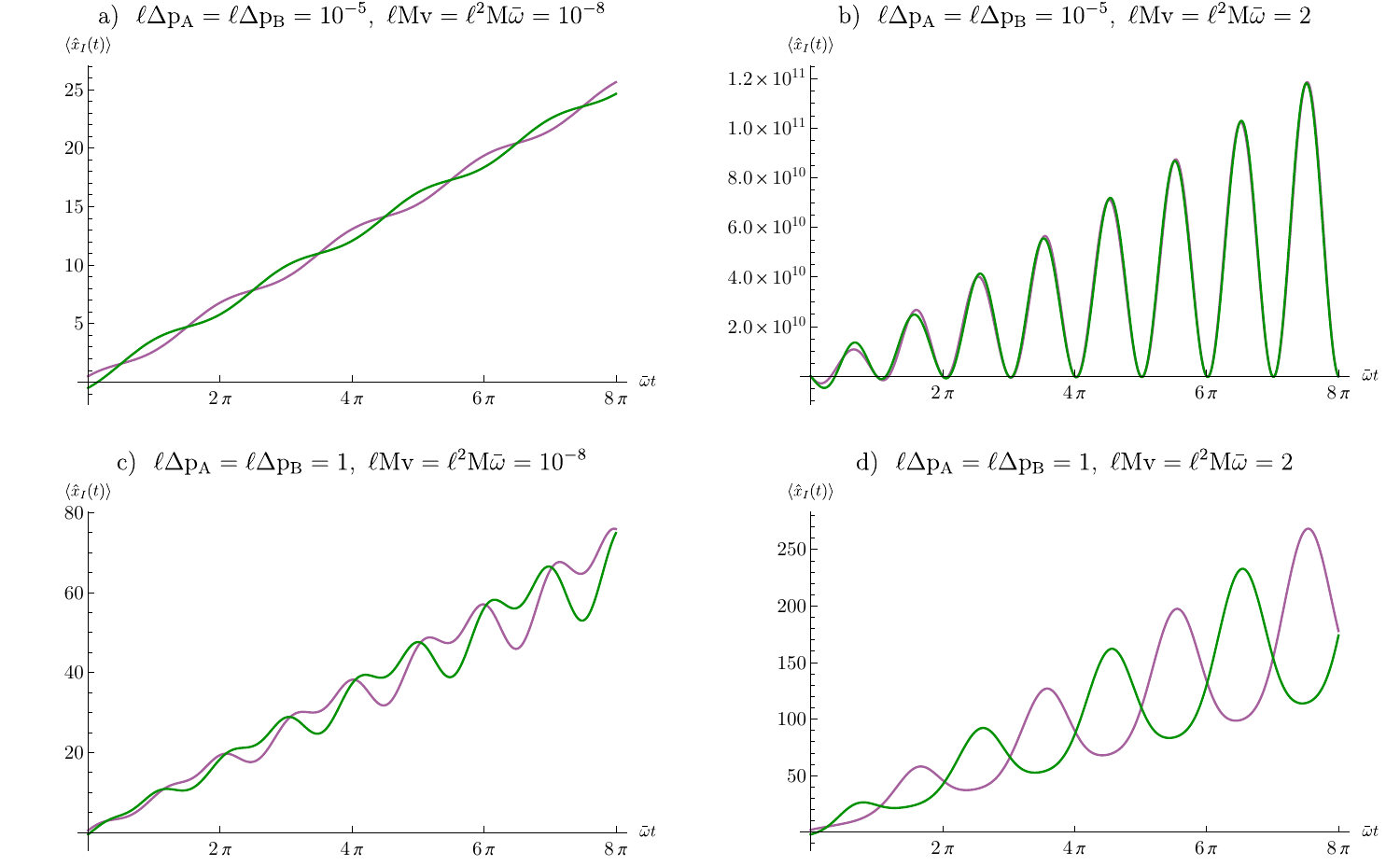}
    \caption{Time evolution of $\braket{\hx_I(t)}$ in units of the coordinate distance between the two particles at $t=0$ with increasing momentum uncertainty from left to right as well as increasing oscillator momentum scale $M\omega$ from top to bottom. The green and violet lines stand for $\braket{\hx_A(t)}$ and $\braket{\hx_B(t)},$ respectively. Additional parameter values are $\braket{\hx_A(0)}=-\braket{\hx_{B}(0)}=\ell/2=1/2$ and $\ell Mv=1.$ The black dots symbolise the end of a period.}
    \label{fig:HarmOscAllboosted}
\end{figure}

As seen from a boosted observer, the evolution is depicted in Fig. \ref{fig:HarmOscAllboosted}. In this case the boost is chosen to be of the order $v t\sim \braket{\hx_I}(0)$ such that the boost evolution does not overpower the harmonic dynamics. As time measured in periods in the plots, \ie $\bar\omega t$, is of order $1,$ this comes down to the relation $M v/M\bar\omega\sim\ell\sim\braket{\hx_I}(0).$ Thus, generally we have $\ell Mv\sim \ell^2M\bar\omega$ as can be seen in the plots. If, then, the boost and system momentum scale, as well as the localisation in position space are small, we recover the ordinary boosted harmonic oscillator (\cf Fig. \ref{fig:HarmOscAllboosted} a). In contrast, at large boosts and system momentum scales, a situation displayed in Fig \ref{fig:HarmOscAllboosted}, both particles start oscillating in phase at much larger distance from the origin, \ie generally $\braket{\hx_I}(t)\gg \braket{\hx_I}(0).$ As demonstrated in Fig \ref{fig:HarmOscAllboosted} c strong localisation in and of itself does not imply significant changes in the boosted with respect to the unboosted case (\cf Fig \ref{fig:HarmOscAllunboosted} c). It is the combination of strong localisation and large masses (Fig \ref{fig:HarmOscAllboosted} d) which is of special interest because it essentially recovers the classical dynamics. The interesting point here lies in the fact that those oscillatory peaks pointing towards the observer (the origin) are softened while those directed away from the observer are sharpened. This property demonstrates experienced by a moving observer. According to Eq. \eqref{eqn:BoostedPosition}, objects in relative motion to the observer in the origin appear more distant depending on their kinetic energy. This effect is stronger for objects which are farther away.   

To summarise, while there are no corrections to the spectrum of the Hamiltonian, the deformation induced by a canonical transformation applied to the ordinary Galilean-invariant Hamiltonian does lead to physical changes because it is the position that we associate the physical interpretation with. If the physical position is given by $\hx_I$ instead of $\hX_I,$ \ie in the presence of a minimal length, the ensuing modifications to the theory are nontrivial.


\section{Conclusion}\label{sec:concl}

A quantum mechanical model with a minimal length requires a cut-off in the eigenspectrum of the wave-number conjugate to the position operator. This implies, on the one hand, that wave numbers can not add up linearly and, on the other hand, that boosts have to act nontrivially on wave numbers. In other words, Galilean relativity has to be either explicitly broken or, at least, deformed in some way.

In this work we have explicitly demonstrated that the only dynamics invariant under deformed Galilean transformations in one dimension is canonically related to ordinary Galilean evolution. In other words, given the position $\hx,$ and its conjugate $\hk,$ we can find another canonical pair $(\hX,\hp)$ in terms of which the Hamiltonian does not appear deformed. The transition from the set $(\hx,\hk)$ to the set $(\hp,\hX)$ is a diffeomorphism in momentum space defined such that the momenta $\hp(\hk)$ compose linearly. In other words, expressed in terms of $\hp,$ the law of conservation of momentum is undeformed. 

Customary minimal-length models, subsumed under the term GUP, purport the existence of a preferred notion of momentum in terms of which the kinetic part of the Hamiltonian is quadratic. Here, introducing the momentum $\hp=p(\hk),$ we have corroborated this assertion, turning the existence of a linearly adding momentum $\hp$ into a necessary condition for the existence of a relativity principle. Indeed, the resulting free-particle Hamiltonian is quadratic in $\hp$. However, contrary to conventional models, deformed Galilean relativity requires the interaction potential between two particles to depend on a generalisation of the respective distance, \ie to be deformed. Therefore, the prevailing GUP models cannot accommodate for a relativity principle.

Semiclassically speaking, the deformation of the boost operator implies that the position of a particle in motion with respect to the observer is modified as a function of its kinetic energy in the observer's frame. For conventional types of models, this change amounts to an increase in distances and apparently elongates extended objects in motion in a way reminiscent of the Lorenz contraction. However, here it is not the relative velocity that is compared to the speed of light but the mass and the kinetic energy that are compared to the minimal-length scale.

That the resulting deformed Galilean-invariant dynamics is canonically related to the ordinary Galilean-relativistic one does not imply that the model is trivial. Indeed, an analogous statement could be made about special relativity in $1+1$ dimensions. While the spectrum of the Hamiltonian is unmodified with respect to the ordinary one, the dynamics of the position operator is clearly deformed. For instance, the study of elastic collisions suggests a revision of the principle of absolute locality in favour of relative locality (for more information see \eg \cite{Amelino-Camelia:2011lvm,Amelino-Camelia:2011hjg}). We have further demonstrated the nontriviality of the dynamics at the instructive example of two particles interacting through a harmonic potential. In particular, a boosted observer does indeed find relative-locality-like effects as displayed in Fig. \ref{fig:HarmOscAllboosted} d.

The apparent triviality of the model is rooted in the fact that a one-dimensional wave-number space cannot be curved. Similarly and in contrast to its higher-dimensional counterparts, the space of velocities in 1+1-dimensional special relativity is flat. By analogy with special and doubly special relativity, we expect this to change in higher dimensions when coordinates cease to commute. Indeed, it has been shown that the curvature of momentum space is proportional to the noncommutativity of the coordinates \cite{Amelino-Camelia:2011lvm,Wagner:2021thc,Wagner:2022dkc,Wagner:2022rjg}. Furthermore, the finding that the existence of a minimal length requires a cut-off in wave-number space generalises to noncommutative geometries \cite{Bosso:2023sxr}. Therefore, it would be interesting to extend the present results to that case. We hope to report back on this matter in the future. 

\appendix

\section{Proof of associative and commutative composition law}\label{app:compLaw}

Equation \eqref{eqn:MomAddProp} constrains the composition laws compatible with any version of deformed Galilean invariance. This appendix is dedicated to analysing this constraint. First, we rewrite Eq. \eqref{eqn:MomAddProp} in terms of $F$ as
\begin{equation}
    \frac{F^{(1,1)}}{F^{(1,0)}F^{(0,1)}}=\tilde{F}(F),\label{eqn:appCond}
\end{equation}
where the superscripts correspond to the number of derivatives with respect to the first and second entries of the function $F(\hk_A,\hk_B),$ respectively. This equation is generally satisfied by a composition law of the kind
\begin{equation}
    F(\hk_A,\hk_B)=p^{-1}\left(p(\hk_A)+p(\hk_B)\right)\label{eqn:pCompLaw}
\end{equation}
for some function $p(k).$ Composition laws of this kind are trivially associative and commutative.

However, for Eq. \eqref{eqn:appCond} to imply a commutative and associative composition law, it is necessary that Eq. \eqref{eqn:pCompLaw} constitutes its unique solution. Here, we demonstrate this by a perturbative analysis to infinite order in $\ell,$ \ie under the assumption that the functions $F,$ $\tilde F$ and $p$ are analytic. In this case the functions, $\tilde F$ and $p,$ being dependent on one variable, have one free coefficient at every order in $\ell.$ The composition law $F$, in turn, is a general function of two momenta thus requiring $n$ different coefficients at order $n.$ Both Eqs. \eqref{eqn:appCond} and \eqref{eqn:pCompLaw} provide $n$ constraints at order $n.$ Thus, both introduce one additional coefficient while providing the same number of constraints on the composition law. As a result, the composition law in both cases has one unconstrained coefficient at every order in $\ell.$ In a nutshell, we have found a solution of Eq. \eqref{eqn:pCompLaw} which does not further constrain the composition law. Thus, we have determined its general solution.

To illustrate how this comes about, we find the said constraints to fourth order in $\ell.$ The wave-number composition can be expanded as
\begin{equation}
    F(\hk_A,\hk_B)=\hk_A+\hk_B+\sum_{n,m=1}^\infty F_{nm} l^{n+m-1}\hk_A^m\hk_B^n.
\end{equation}
Furthermore, bearing in mind that it has to have dimensions of length, we may express the function $\tilde{F}(\hk)$ as
\begin{equation}
    \tilde{F}(\hk)=\ell \sum_{n=0}^\infty \tilde{F}_n(\ell\hk)^n.
\end{equation}
As a result, we can expand Eq. \eqref{eqn:appCond} in powers of $\ell$ and compare the coefficients of powers of $\hk_A,$ $\hk_B$ to obtain constraints on a given composition law and determine the corresponding $\tilde{F}_n.$ As the present appendix is centred around the composition law, we only display the former. To fourth order in $\ell,$ they read
\begin{align}
    F_{1,2}=&F_{2,1},\qquad F_{1,3}=F_{3,1}\qquad F_{2,2}=\frac{3 F_{3,1}}{2}+F_{1,1} F_{2,1},\\
    F_{1,4}=&F_{4,1}, \qquad F_{2,3}=F_{3,2}=\frac{1}{2} \left(2 F_{2,1}^2+3 F_{1,1} F_{3,1}\right)+2 F_{4,1}.
\end{align}
Indeed, there is one free coefficient at every order (\ie $F_{1,1},$ $F_{2,1},$ $F_{3,1}$ and $F_{4,1}$). Thus, at that order the composition law becomes
\begin{align}
    F=&\hk_A+\hk_B+F_{1,1} \hk_A \hk_B\ell+ F_{2,1} \hk_A \hk_B \left(\hk_A+\hk_B\right)\ell^2+\frac{1}{2} \hk_A \hk_B \left[F_{3,1} \left(3 \hk_A \hk_B+2 \hk_A^2+2 \hk_B^2\right)+2 F_{1,1} F_{2,1} \hk_A \hk_B\right]\ell^3\nonumber\\
    &+\frac{1}{2} \hk_A \hk_B \left(\hk_A+\hk_B\right) \left[2 F_{2,1}^2 \hk_A \hk_B+3 F_{1,1} F_{3,1} \hk_A \hk_B+2 F_{4,1} \left(\hk_A \hk_B+\hk_A^2+\hk_B^2\right)\right]\ell^4.
\end{align}
This function is clearly invariant under the exchange $A\leftrightarrow B,$ \ie the composition law is commutative. Furthermore, it can be explicitly shown that $F(\hk_{A},F(\hk_{B},\hk_C))=F(F(\hk_{A},\hk_B),\hk_{C}),$ which amounts to associativity. Indeed, both Eqs. \eqref{eqn:appCond} and \eqref{eqn:pCompLaw} imply the same composition law in the same parameterisation.

\providecommand{\href}[2]{#2}\begingroup\raggedright\endgroup

\end{document}